 \documentclass[final,3p,times, 12pt]{elsarticle}

\usepackage{amssymb}
\usepackage{graphicx}

\usepackage{amsmath}
\usepackage{amssymb}
\usepackage{caption}
\usepackage{booktabs}
\usepackage{tabls}
\usepackage{multirow}
\usepackage{listings}

\usepackage{color}

\definecolor{OliveGreen}{cmyk}{0.64,0,0.95,0.40}
\definecolor{colFunc}{rgb}{1,0.07,0.54}
\definecolor{CadetBlue}{cmyk}{0.62,0.57,0.23,0}
\definecolor{Brown}{cmyk}{0,0.81,1,0.60}
\definecolor{colID}{rgb}{0.63,0.44,0}
 
\def\imp{\operatorname{\mathfrak{Im\,}}}
\def\rep{\operatorname{\mathfrak{Re\,}}}

\newcommand{\fourier}[1]{\mathcal{F} \biggl[#1 \biggl]}
\newcommand{\ifourier}[1]{\mathcal{F}^{-1} \biggl[#1 \biggl]}
\newcommand{\sfourier}[1]{\mathcal{F} \bigl[#1 \bigr]}
\newcommand{\sifourier}[1]{\mathcal{F}^{-1} \bigl[#1 \bigr]}

\journal{Computer Physics Communications}

\begin{document}

\begin{frontmatter}



\title{Fast computation of computer-generated hologram using\\
Xeon Phi coprocessor}


\author[chiba]{Koki Murano}
\author[chiba]{Tomoyoshi Shimobaba\corref{cor1}}
\ead{shimobaba@faculty.chiba-u.jp}
\author[chiba]{Atsushi Sugiyama}
\author[kochi]{Naoki Takada}
\author[chiba]{Takashi Kakue}
\author[chiba]{Minoru Oikawa}
\author[chiba]{Tomoyoshi Ito}

\cortext[cor1]{Corresponding author}
\address[chiba]{Graduate School of Engineering, Chiba University, 1--33 Yayoi--cho, Inage--ku, Chiba 263--8522, Japan}
\address[kochi]{Faculty of Science, Kochi University, 2--5--1, Akebono--cho, Kochi 780--8520, Japan}

\begin{abstract}
We report fast computation of computer-generated holograms (CGHs) using Xeon Phi coprocessors, which have massively x86-based processors on one chip, recently released by Intel.
CGHs can generate arbitrary light wavefronts, and therefore, are promising technology for many applications: for example, three-dimensional displays, diffractive optical elements, and the generation of arbitrary beams.
CGHs incur enormous computational cost.
In this paper, we describe the implementations of several CGH generating algorithms on the Xeon Phi, and the comparisons in terms of the performance and the ease of programming between the Xeon Phi, a CPU and graphics processing unit (GPU).
\end{abstract}


\end{frontmatter}


\section{Introduction}
Holography \cite{gabor} is used to record light waves including the amplitude and phase on a photographic plate as interference patterns, as a ``hologram,'' and reconstruct the light wave from the hologram, utilizing its interference and diffraction phenomena.
To simulate light propagation on a computer, we can numerically generate computer-generated holograms (CGHs). 
CGHs can generate arbitrary light waves, and therefore, are promising technology for many applications: for example, three-dimensional displays \cite{benton, slinger}, diffractive optical elements \cite{doe}, and the generation of arbitrary beams \cite{beam}.

If we apply CGHs to a 3D display, the technique becomes very attractive because it is difficult for other 3D display techniques to reconstruct light waves of original 3D objects prepared by computer graphics or acquired by 3D cameras such as Kinect.
In contrast, CGH can faithfully reconstruct the light waves.
However, the computational time of CGH required for the light propagation from a 3D object hampers the realization of a practical CGH-based 3D display.

Many algorithms have been proposed for accelerating the computational time.
These algorithms are mainly categorized by how to express a 3D object by point cloud and polygon expressions.
Point cloud expression is expressed by 3D objects as the aggregation of self-illuminated points.
Look-up table (LUT) methods \cite{lucente, nlut} store pre-computed light waves on a CGH, and then, accumulate light waves to generate a final CGH pattern.
Recurrence relation methods \cite{recur1,recur2,recur3} calculate light waves on a CGH using recurrence relations, instead of directly calculating the optical path that is time-consuming.   
To place a 3D object near a CGH, the image hologram method \cite{image_hol} accelerates the calculation of an image CGH.  
Te wavefront recording method \cite{wr1,wr2,wr3,wr4} accelerates the calculation of a CGH to introduce a virtual plane between a 3D object and the CGH.  
It consists of two steps: the first step is the calculation of the virtual plane from the 3D object. 
In the second step, we obtain the CGH to execute diffraction calculation from the virtual plane.
Polygon methods have been vigorously researched \cite{poly1,poly2,poly3,poly4,poly5}. 
The key to the polygon method is how to calculate tilted polygons with regards to a CGH because normal diffraction calculation only calculate parallel planes.
Other methods are, for example, band-limited double step Fresnel diffraction \cite{bldsf} to accelerate diffraction calculation, the Fresnel integral method \cite{freapprox} that is faster than the methods using fast Fourier transform (FFT), the efficient calculation of color CGHs using color space conversion \cite{color} and so forth.

Using hardware is an effective means to further boost computational speed for CGH because the pixels on a CGH can be calculated independently; therefore the CGH calculation is very suitable for hardware implementation. 
We developed special-purpose computers for CGH to dramatically increase the computational speed \cite{horn1,horn2,horn3,horn4,horn5,horn6}.
This is called ``HOlographic ReconstructioN (HORN)''. 
The HORN computers designed by pipeline architecture can calculate light waves on a CGH at high speed. 
The HORN computers were implemented on a field programmable gate array (FPGA) board, except HORN-1 and -2. 
To date, we have developed six HORN computers.
Another major hardware is the graphics processing unit (GPU).
Recent GPUs allow us to use a highly parallel processor, because the GPUs have many simple processors that can execute floating-point arithmetic operations. 
GPUs were widely used in the realm of CGH \cite{gpu1,gpu2,gpu3,gpu4}.
Although GPUs have merits in terms of computational power and the low cost of the GPU board, we need to use special programming languages (e.g. CUDA, OpenCL) to program on GPUs.

In this paper, we report fast computation of CGH using Xeon Phi co-processors.
Intel very recently launched the Xeon Phi Coprocessor \cite{phi} that has massive x86-based processors on one chip.
One of the merits of Xeon Phi is that we can program on the Xeon Phi using standard C and various libraries for parallel environments (e.g. OpenMP, MPI). 
Therefore, we gain much acceleration of CGH calculation with only a small amount of labor, compared with GPUs.
In Section 2, we explain our CGH algorithms which are implemented on a CPU, GPU and Xeon Phi: recurrence relation method and wave-front recording method.
In Sections 3 and 4,  we describe the outline of the Xeon Phi co-processor and the implementations of the CGH algorithms for each processor.
Section 5 shows the comparison of the performance and the ease of  programming between the Xeon Phi, a CPU and GPU.
In Section 6, we conclude this work.

\section{CGH algorithms}

We describe three algorithms for CGH calculation in point cloud expression: direct calculation, recurrence relation and wavefront recording methods.
CGH can be calculated from a 3D object data generated by a 3D modeling software or actual 3D data acquired by a 3D camera such as Kinect.
We first describe the direct calculation.  

\subsection{Direct calculation}
Let us suppose that a virtual 3D object consists of $N$ point light sources.
The coordinate for $j$-th point light source is expressed as $(x_j, y_j, z_j)$.
The distribution of an object light $O(x_h,~y_h)$ on a CGH can be calculated by accumulating light propagation 
from all of the point light sources, that is,
\begin{eqnarray}
O(x_h,~y_h)&=&\sum_j^N A_j {\rm exp}(j \frac{2 \pi}{\lambda} \sqrt{(x_h - x_j)^2 + (y_h-y_j)^2 + z_j^2 }),  \\
&\approx& \sum_j^N {A_j}{\rm exp}(\frac{2 \pi}{\lambda} (z_j + \frac{(x_h-x_j)^2+(y_h-y_j)^2}{2 z_j})),
\label{eqn:eq_object}
\end{eqnarray}
where, $\lambda$ is the wavelength of a reference light, and $A_j$ is the amplitude of the $j$-th point light source.
The approximation of the distance calculation in Eq.(\ref{eqn:eq_object}) is Fresnel approximation.
We call these equations {\it direct calculation}, because the calculation is the direct summation of object lights.

CGH types are mainly categorized into amplitude or phase only CGH (a.k.a. Kinoform).
We can calculate a phase only CGH by taking the phase from the complex amplitude of Eq.(1) or (2),
\begin{equation}
\theta(x_h,~y_h)={\rm tan^{-1}}\left(\frac{\imp \left\{O(x_h,~y_h) \right\}}{\rep\left\{O(x_h,~y_h) \right\}}\right),
\label{eqn:eq_phase}
\end{equation}
where the operators $\rep\left\{\cdot \right\}$ and $\imp \left\{\cdot \right\}$ take the real and imaginary parts of the complex number.

We calculate amplitude CGHs by the interference between an object light and reference light $R(x_h,~y_h)=A_{r} \exp(j \phi_r(x_h,  y_h))$, 
where $A_r$ and $\phi_r(x_h, y_h)$ are the amplitude and phase of the reference light.
The intensity of the interference is expressed as,
\begin{eqnarray}
I(x_h,~y_h)&=&|O(x_h, y_h) + R(x_h,y_h)|^2 \\ \nonumber
&=& |O(x_h,y_h)|^2 + |R(x_h,y_h)|^2 + 
 O(x_h,y_h) R^{*}(x_h,y_h) + O^{*}(x_h,y_h) R(x_h,y_h),
\label{eqn:eq_intensity}
\end{eqnarray}
where, * denotes complex conjugate.
In amplitude CGH calculation, we can omit the first and second terms because these terms do not include the 3D information of a 3D object.
Therefore, we can rewrite the above equation as follows:
\begin{eqnarray}
I(x_h,~y_h) &=& O(x_h,y_h) R^{*}(x_h,y_h) +  O^{*}(x_h,y_h) R(x_h, y_h) \\ \nonumber
&=& 2{A_r} \sum_j^N {A_j}{\rm cos}(\frac{2 \pi}{\lambda} \sqrt{(x_h -  x_j)^2 + ( y_h- y_j)^2 + z_j^2 } + \phi_r), \\ \nonumber
&\approx& \sum_j^N {A_j}{\rm cos}(\frac{2 \pi}{\lambda} (z_j + \frac{(x_h-x_j)^2+(y_h-y_j)^2}{2 z_j}) + \phi_r),
\label{eqn:eq_cgh_basic}
\end{eqnarray}
where, the constant value $2A_r$ can be omitted because it is not important to reconstruct the 3D object.

\subsection{Recurrence relation method}

Our recurrence relation method \cite{recur3, horn4} can compute the phase components of the exponential function in Eq.(\ref{eqn:eq_object}) by the following two recurrence relations:
\begin{eqnarray}
\Gamma_n&=&\Gamma_{n-1}+\delta_{n-1}, \\
\delta_{n}&=&\delta_{n-1}+\Delta, 
\label{eqn:recur1}
\end{eqnarray}
where $\Gamma_1 = P_j((x_h - x_j)^2 + (y_h - y_j)^2), \delta_1= P_j (2(x_h - x_j) + 1), \Delta=2 P_j$ and $P_j=1 / (2 \lambda z_j)$.
Using the recurrence relations, the object waves on the CGH can be calculated by the following equation: 
\begin{equation}
O(x_h + n,~y_h)=\sum_j^N {A_j}{\rm exp} (2\pi(\Gamma_n + \delta_n)).
\label{eqn:recur2}
\end{equation}

Namely, after calculating the phase at $(x_h, y_h)$, we can obtain the adjacent phase at $(x_h+n, y_h)$ by Eqs.(6) and (7).
The calculation cost can be reduced, compared with the direct calculation of Eq.(\ref{eqn:eq_object}). 

\subsection{Wavefront recording method}

The wavefront recording method reduces the computational complexity of CGH by two-step calculations.
In the first step of the wavefront recording method, we introduce 
virtual plane $u_w(x_w,y_w)$ between a 3D object and CGH, and record the complex amplitude from the 3D object using Eq.(1) or (2) on the virtual plane.

When placing the virtual plane near the 3D object, the object light traverses small regions  on the virtual plane.
Therefore, the computational amount of Eq.(1) or (2) can be reduced, compared with the case where the virtual plane is not introduced.

The second step calculates the complex amplitude of 3D object light $O(x_h,y_h)$ on the CGH using the diffraction calculation between the virtual plane and CGH.
Several diffraction calculations were proposed.
For example, Fresnel diffraction which is a well-known diffraction calculation, is as follows:
\begin{eqnarray}
O(x_h,y_h) &=& 
\frac{\exp(i \frac{2 \pi}{\lambda} z)}{i \lambda z}  \int \!\!\int u_w(x_w, y_w)  \exp(i \frac{\pi}{\lambda z} ((x_h-x_w)^2+(y_h-y_w)^2) ) dx_w dy_w  \\
&=& \frac{\exp(i \frac{2 \pi}{\lambda} z)}{i \lambda z}  \ifourier{ \fourier{u_w(x,y)} \cdot \fourier{h(x,y)} } 
\label{eqn:fresnel}
\end{eqnarray}
where, $\sfourier{\cdot}$ and $\sifourier{\cdot}$ are the Fourier and inverse Fourier operators, $z$ is the distance between the virtual plane and the CGH, and the impulse response $h(x,y)= \exp(\frac{i \pi}{\lambda z} (x^2+y^2))$.
Of course, the Fourier transform can be accelerated by FFT. 
Lastly, we calculate the phase only or amplitude CGH by Eq.(\ref{eqn:eq_phase}) or (\ref{eqn:eq_intensity}).

\section{Xeon Phi coprocessor}
\begin{figure}[h]
	\begin{center}
		\includegraphics[width=100mm]{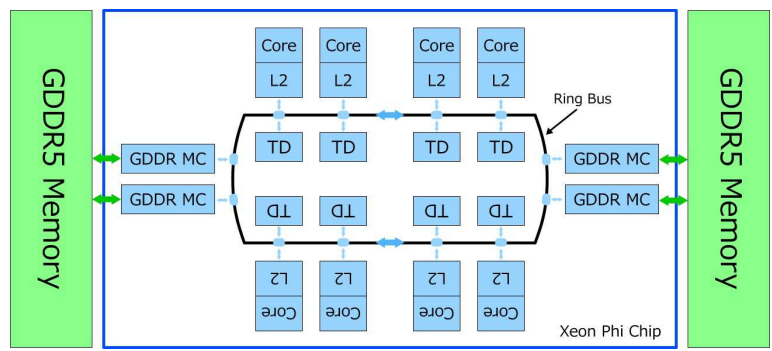}
 	\end{center}
	\caption{Conceptual diagram of Xeon Phi Coprocessor.}
	\label{fig:diagram}
\end{figure}

The Xeon Phi coprocessor has many cores based on the concepts of the Intel x86 architecture and offers a standard shared-memory architecture.
In this work, we used Xeon Phi 5110P. Table.\ref{tab:spec} shows the specification of Xeon Phi 5110P.
As shown in Table.\ref{tab:spec}, the Xeon Phi coprocessor has 60 cores and every core offers four-way simultaneous multi-threading (SMT). 
Therefore, the total number of threads is 240. 
In addition, every core has single instruction multiple data (SIMD) vectors with 512-bit wide, which can address the arithmetic operations of eight double-precision or sixteen single-precision floating point numbers simultaneously.
Figure \ref{fig:diagram} shows the conceptual diagram of the Xeon Phi coprocessor. 
The Xeon Phi has a structure in which a bidirectional ring bus connects each core. 
By the ring bus, each core can access a GDDR5 memory through GDDR memory controller (GDDR MC) and other cores. 
Each core has a private L2 cache kept fully coherent by a tag directory (TD).
The Xeon Phi coprocessor is connected via a PCI Express bus to a host system.

\begin{table}[h]
	\caption{Specification of Xeon Phi 5110P.}
	\begin{center}
	\begin{tabular}{c|c} \hline
	Core & 60 \\
	Thread & 240 \\ 
	L1 cache & 32[kB/core] \\
	L2 cache & 512[kB/core] \\
	Clock & 1.053 [GHz] \\
	Memory & 8 [GB] \\
	Memory Band Width & 320 [GB/s] \\ \hline
	\end{tabular}
	\end{center}
	\label{tab:spec}
\end{table}

The Xeon Phi coprocessor can be programmed in two different ways. The former is ``native mode'' that runs a whole program on the coprocessor.
When using the native mode, the coprocessor looks like one node PC connected to the host PC. 
The host PC transfers the executable file for the coprocessor and runs it on the coprocessor via SSH (Secure SHell).
The native mode has the disadvantage that it has slow running time for a sequential part in the program because the clock frequency of the core of the Xeon Phi coprocessor is slower than a modern CPU. In addition, the core design is simpler than a modern CPU, so that the core cannot process instructions by out-of-order execution.

The latter is ``offload mode,'' which only runs user-specified parts in the program on the coprocessor. When using the offload mode, the coprocessor looks like accelerators (e.g. GPUs).
It is desirable that the specified parts have high parallelism and huge computational burden if we induce the high performance of the coprocessor. 
The user-specified parts are specified by pragma directives.
In this work, we used the offload mode because the CGH calculation part in a whole program has extremely high parallelism and huge computational burden.

\section{Implementation of CGH algorithms on Xeon Phi coprocessor}
In this section, we show the actual source codes for CGH algorithms as described in Section 2.
In addition, in terms of the ease of programming, we compare source codes on a CPU, Xeon Phi and GPU, respectively.  
One of the merits for the usage of Xeon Phi is the ease of the programming, compared with GPUs.

\subsection{Direct calculation}
List \ref{lst:direct_CPU} shows a part of the source code for phase only CGHs using the direct calculation of Eq.(\ref{eqn:eq_object}) and Eq.(\ref{eqn:eq_phase}) on a general multi-core CPU by standard C and OpenMP.
In the source code, the most outer loop of the CGH calculation is automatically divided and assigned to each CPU thread by the pragma directive (\#pragma omp parallel for).

\begin{lstlisting}[caption={Source code on CPU (direct calculation).}, label=lst:direct_CPU, language=C, numbers=left, numberstyle=\tiny, stepnumber=1, frame=single]
#define PI					3.1415926535F	 
#define P					8.0e-6					//CGH sampling pitch (8um)
#define WL				633e-9F				//wave length (633nm:red)
#define NORM_CNT	40.5845104884		//255/(2*PI)
#define NUMBER		10000					//number of object points
#define WIDTH			1920						//width of CGH
#define HEIGHT		1080						//height of CGH

//static memory
unsigned char holo[WIDTH*HEIGHT];

int main(void){

	//dynamic memory
	float *obj_x,*obj_y,*obj_z;

	obj_x = (float*)malloc(sizeof(float)*NUMBER);
	obj_y = (float*)malloc(sizeof(float)*NUMBER);
	obj_z = (float*)malloc(sizeof(float)*NUMBER);
	
	//***************
	//Input data
	//***************

	//following CGH calculation part
	#pragma omp parallel for									//pragma for OpenMP
	for(int i=0;i<HEIGHT;i++){
		for(int j=0;j<WIDTH;j++){
			unsigned int adr = j + WIDTH * i;
			float re = 0.0f;
			float im = 0.0f;
			for(int k=0;k<NUMBER;k++){
				float x = P*j - obj_x[k];
				float y = P*i - obj_y[k];
				float theta = PI*WL*(x*x+y*y)/obj_z[k];	//phase term
				re += cosf(theta);								//accumulate real part
				im += sinf(theta);								//accumulate imagnary part
			}
			float psi = atan2f(im,re);							//get argument
			if(psi < 0.0f)	psi += 2*PI;
			holo[adr] = (unsigned char)(NORM_CNT*psi);	//normalization
		}
	}
}
\end{lstlisting}

And, List \ref{lst:direct_Phi} shows the source code on the Xeon Phi coprocessor by the offload mode.
The difference between Lists \ref{lst:direct_CPU} and \ref{lst:direct_Phi} is the only offload pragma directives.
As entioned below, the Xeon Phi programming becomes easy compared with GPU programming because GPU programming requires writing of kernel functions and its invoking functions, memory allocation and transfer functions using CUDA APIs.

In line 10 of List \ref{lst:direct_Phi}, the directive ``\_\_attribute\_\_(target(mic:0))" represents the allocation of the static memory on both the Xeon Phi coprocessor and host PC. 
In line 26-30 of List \ref{lst:direct_Phi}, the pragma directive ``\#pragma offload target(mic:0)" indicates that the part surrounded by the pragma directive is run on the coprocessor, which is offload mode.

If the host PC has some Xeon Phi coprocessors, we need to select the coprocessors by ``target" keyword.
In line 27, ``in(obj\_x:length(NUMBER))" is a keyword for transferring the input data from the host PC to the coprocessor. 
The modifier ``length()" indicates the number of elements for dynamic memory. 
We do not need it for static memory.
Similarly, in line 30, ``out(holo)" is a keyword for transferring the output data from the coprocessor to the host PC. 
Usually, the memory allocation is performed implicitly on the coprocessor.

\begin{lstlisting}[caption={Source code on Xeon Phi coprocessor (direct calculation).}, label=lst:direct_Phi, language=C, numbers=left, numberstyle=\tiny, stepnumber=1, frame=single]
#define PI			3.1415926535F	
#define P			8.0e-6F			//CGH sampling pitch (8um)
#define WL			633e-9F			//wave length (633nm:red)
#define NORM_CNT	40.5845104884F	//255/(2*PI)
#define NUMBER		10000			//number of object points
#define WIDTH		1920			//width of CGH
#define HEIGHT		1080			//height of CGH

//static memory
__attribute__((target(mic:0))) unsigned char holo[WIDTH*HEIGHT];

int main(void){

	//dynamic memory
	float *obj_x,*obj_y,*obj_z;
 
	obj_x = (float*)malloc(sizeof(float)*NUMBER);
	obj_y = (float*)malloc(sizeof(float)*NUMBER);
	obj_z = (float*)malloc(sizeof(float)*NUMBER);

	//***************
	//Input data
	//***************

	//following CGH calculation part
	#pragma offload target(mic:0) \
		in(obj_x:length(NUMBER))\
		in(obj_y:length(NUMBER))\
		in(obj_z:length(NUMBER))\
		out(holo)
	{
		#pragma omp parallel for				//pragma for OpenMP
		for(int i=0;i<HEIGHT;i++){
			for(int j=0;j<WIDTH;j++){
				unsigned int adr = j + WIDTH * i;
				float re = 0.0f;
				float im = 0.0f;
				for(int k=0;k<NUMBER;k++){
					float x = P*j - obj_x[k];
					float y = P*i - obj_y[k];
					float theta = PI*WL*(x*x+y*y)/obj_z[k];		//phase term
					re += cosf(theta);								//accumulate real part
					im += sinf(theta);								//accumulate imaginary part
				}
				float psi = atan2f(im,re);							//get argument
				if(psi < 0.0f)	psi += 2*PI;
				holo[adr] = (unsigned char)(NORM_CNT*psi);	//normalization
			}
		}
	}
}
\end{lstlisting}

For comparison, in List 3, we show the source code on a GPU by CUDA, corresponding to List 1. 
In line 13-28 of List \ref{lst:direct_GPU}, ``\_\_global\_\_ void cgh\_calc" function is a GPU kernel. 
When using a GPU, it is necessary to explicitly rewrite the computational part as a GPU kernel, unlike Xeon Phi coprocessors. 
In addition, in order to transfer the data between the host PC and GPU,  and allocate the device memory, we must write them explicitly by using CUDA APIs of ``cudaMalloc'' and ``cudaMemcpy'' (lines 48-57).
Furthermore, we need to invoke the kernel function using CUDA API, as shown in line 60 of List 3.

\begin{lstlisting}[caption={Source code on GPU by CUDA (direct calculation).}, label=lst:direct_GPU, language=C, numbers=left, numberstyle=\tiny, stepnumber=1, frame=single]
#define PI			3.1415926535F	
#define P			8.0e-6F			//CGH sampling pitch (8um)
#define WL			633e-9F			//wave length (633nm:red)
#define NORM_CNT	40.5845104884F	//255/(2*PI)
#define NUMBER		10000			//number of object points
#define WIDTH		1920			//width of CGH
#define HEIGHT		1080			//height of CGH

#define TX			32				//x direction thread per blocks
#define TY			18				//y direction thread per blocks

//GPU kernel
__global__ void cgh_calc(float *o_x, float *o_y, float* o_z, unsigned char* h){
	int tx = threadIdx.x + blockDim.x * blockIdx.x;
	int ty = threadIdx.y + blockDim.y * blockIdx.y;
	float re = 0.0f;
	float im = 0.0f;
	for(int k=0;k<NUMBER;k++){
		float x = P*tx - obj_x[k];
		float y = P*ty - obj_y[k];
		float theta = PI*WL*(x*x+y*y)/obj_z[k];
		re += cosf(theta);
		im += sinf(theta);
	}
	float psi = atan2f(im,re);
	if(psi < 0.0f) psi += 2*PI;
	h[ty*WIDTH+tx] = (unsigned char)(NORM_CNT*psi);
}

int main(void){

	float *obj_x,*obj_y,*obj_z;

	obj_x = (float*)malloc(sizeof(float)*NUMBER);
	obj_y = (float*)malloc(sizeof(float)*NUMBER);
	obj_z = (float*)malloc(sizeof(float)*NUMBER);

	//****************
	//data input etc...
	//****************
	
	dim3 grids(WIDTH/TX,HEIGHT/TY);		//grid size
	dim3 threads(TX,TY);				//block size
	
	float *d_x,*d_y,*d_z;
	unsigned char *d_h;
	
	//GPU memory allocation
	cudaMalloc((void**)&d_x,sizeof(float)*NUMBER);
	cudaMalloc((void**)&d_y,sizeof(float)*NUMBER);
	cudaMalloc((void**)&d_z,sizeof(float)*NUMBER);
	cudaMalloc((void**)&d_h,sizeof(unsigned char)*WIDTH*HEIGHT);

	//trasfer input data from host to GPU
	cudaMemcpy(d_x,obj_x,sizeof(float)*NUMBER,cudaMemcpyHostToDevice);
	cudaMemcpy(d_y,obj_y,sizeof(float)*NUMBER,cudaMemcpyHostToDevice);	
	cudaMemcpy(d_z,obj_z,sizeof(float)*NUMBER,cudaMemcpyHostToDevice);

	//kernel execution
	cgh_calc<<<grids,threads>>>(d_x,d_y,d_z,d_h);

	//trasfer result data from GPU to host
	cudaMemcpy(holo,d_h,sizeof(unsigned char)*WIDTH*HEIGHT,cudaMemcpyDeviceToHost);
}
\end{lstlisting}

In regards to the recurrence relation method, we can implement the CGH computation on the Xeon Phi coprocessor in the same way as in List \ref{lst:direct_Phi}.
\subsection{Fresnel diffraction}
The CGH calculation using the wavefront recording method has two steps.
The implementation of the first step is almost the same as the direct calculation. 
The second step is the Fresnel diffraction expressed by Eq.(\ref{eqn:fresnel}).
List \ref{lst:fresnel_CPU} shows the Fresnel diffraction on the CPU. 
The Fresnel diffraction consists of three FFTs, the generation of the impulse response and one complex multiplication. 
We used Intel Math Kernel Library (MKL) as the FFT library.
Note that, during the calculation of Fresnel diffraction, we have to expand the calculation size by zero-padding to $2N \times 2N$, where $N$ is the horizontal or vertical pixel number, in order to avoid wraparound by the circular convolution of Eq.(\ref{eqn:fresnel}).
After finishing the calculation of  Eq.(\ref{eqn:fresnel}), we crop the wanted area with $N \times N$.

List \ref{lst:fresnel_Phi} shows the implementation on the Xeon Phi coprocessor. 
As well as List \ref{lst:direct_Phi}, by using pragma directives, we can rewrite the source code for the Xeon Phi coprocessor by minimum labor, from List \ref{lst:fresnel_CPU}.
%

\begin{lstlisting}[caption={Source code on CPU (Fresnel diffraction).}, label=lst:fresnel_CPU, language=C, numbers=left, numberstyle=\tiny, stepnumber=1, frame=single]
#define N	1024		//size N*N of input plane
#define N2	2048		//2*N

#define P	8.0e-6F		//sampling pitch
#define WL	633e-9		//wavelength
#define PI	3.1415926535F

#define D	1.0F		//propagated distance

int main(void)
{
	MKL_Complex8 *pln;		//source and destination plane array
	MKL_Complex8 *data;		//data(expanded source and destination area) array
	MKL_Complex8 *prop;		//propagation term array
	
	//DFTI handler
	 DFTI_DESCRIPTOR_HANDLE hand0 = 0;
	 DFTI_DESCRIPTOR_HANDLE hand1 = 0;

	//memory allocation
	pln	 = (MKL_Complex8*)malloc(N*N*sizeof(MKL_Complex8);
	data = (MKL_Complex8*)malloc(N2*N2*sizeof(MKL_Complex8));
	prop = (MKL_Complex8*)malloc(N2*N2*sizeof(MKL_Complex8));
	
	//data input
	DataInput(pln,"input.bmp");

	//the area of the source and destination planes must be expanded from N*N to N2*N2
	#pragma omp parallel for
	for(int j=0;j<N2;j++){
		for(int i=0;i<N2;i++){
			int d_idx = i + N2 * j;
			if(i >= N/2 && j >= N/2 && i < N/2*3 && j < N/2*3){
				int p_idx =  (i-N/2) + (j-N/2) * N;
				data[d_idx] = pln[p_idx];
			}else{
				data[data_idx].real = 0.0f;
				data[data_idx].imag = 0.0f;
			}
		}
	}

	//Transfer function
	#pragma omp parallel for
	for(int j=0;j<N2;j++){
		for(int i=0;i<N2;i++){
			unsigned int idx = i + N2 * j;
			float dx;
			float dy;

			if(x < N && y < N){
				dx = (float)(i);
				dy = (float)(j);
			}else if(x >= N && y < N){
				dx = (float)(x - N2);
				dy = (float)(y);
			}else if(x < N && y >= N){
				dx = (float)(x);
				dy = (float)(y - N2);
			}else{
				dx = (float)(x - N2);
				dy = (float)(y - N2);
			}

			float th = PI*P*(x*x+y*y)/(D*WL);
			prop.real = cosf(th);
			prop.imag = sinf(th);
		}
	}

	//create and initialize DFTI descriptor
	MKL_LONG NN[2];	NN[0] = N2;	NN[1] = N2;
	//create DFTI descriptor
	DftiCreateDescriptor(&hand0 ,DFTI_SINGLE ,DFTI_COMPLEX, 2, NN);
	DftiCreateDescriptor(&hand1 ,DFTI_SINGLE ,DFTI_COMPLEX, 2, NN);
	//initialize
	DftiCommitDescriptor(hand0);
	DftiCommitDescriptor(hand1);
	
	//data plane and propagation term FFT Forward
	DftiComputeForward(hand0, data);
	DftiComputeForward(hand1, prop);

	//complex multiplication
	#pragma omp parallel for
	for(int i=0;i<N2*N2;i++){
		float dre = data[i].real;
		float dim = data[i].imag;
		float pre = prop[i].real;
		float pim = prop[i].imag;
		data[i].real = dre*pre-dim*pim;
		data[i].imag = dre*pim+dim*pre;
	}
	
	//FFT Backward
	DftiComputeBackward(hand0,data);
	
	//data cripping
	#pragma omp parallel for
	for(int j=0;j<N;j++){
		for(int i=0;i<N;i++){
			int d_idx = (i + N/2) + (j + N/2) * N2;
			int p_idx = i + j * N;
			pln[idx] = d[idx];
		}
	}

	//descriptor free
	DftiFreeDescriptor(&hand0);
	DftiFreeDescriptor(&hand1);
}
\end{lstlisting}

\begin{lstlisting}[caption={Source code on Xeon Phi coprocessor (Fresnel diffraction).}, label=lst:fresnel_Phi, language=C, numbers=left, numberstyle=\tiny, stepnumber=1, frame=single]
#define N	1024		//size N*N of input plane
#define N2	2048		//2*N

#define P	8.0e-6F		//sampling pitch
#define WL	633e-9		//wavelength
#define PI	3.1415926535F

#define D	1.0F		//propagated distance

int main(void)
{
	__attribute__((target(mic:0))) MKL_Complex8 *pln;		//source and destination plane array
	__attribute__((target(mic:0))) MKL_Complex8 *data;		//data(expanded source and destination area) array
	__attribute__((target(mic:0))) MKL_Complex8 *prop;		//propagation term array
	
	//DFTI handler
	 __attribute__((target(mic:0))) DFTI_DESCRIPTOR_HANDLE hand0 = 0;
	 __attribute__((target(mic:0))) DFTI_DESCRIPTOR_HANDLE hand1 = 0;

	//host memory allocation
	pln	 = (MKL_Complex8*)malloc(N*N*sizeof(MKL_Complex8);
	data = (MKL_Complex8*)malloc(N2*N2*sizeof(MKL_Complex8));
	prop = (MKL_Complex8*)malloc(N2*N2*sizeof(MKL_Complex8));
	
	//data input
	DataInput(pln,"input.bmp");

	//memory allocation and data transfer from host to Xeon Phi
	#pragma offload target(mic:0)\
		in(pln : length(N*N) alloc_if(1) free_if(0))\
		nocopy(data : length(N2*N2) alloc_if(1) free_if(0))\
		nocopy(prop : length(N2*N2) alloc_if(1) free_if(0))
	{
	}

	#pragma offload(target(mic:0))\
		nocopy(pln : length(N*N) alloc_if(0) free_if(0))\
		nocopy(data : length(N2*N2) alloc_if(0) free_if(0))\
	{
		//the area of the source and destination planes must be expanded from N*N to N2*N2
		#pragma omp parallel for
		for(int j=0;j<N2;j++){
			for(int i=0;i<N2;i++){
				int d_idx = i + N2 * j;
				if(i >= N/2 && j >= N/2 && i < N/2*3 && j < N/2*3){
					int p_idx =  (i-N/2) + (j-N/2) * N;
					data[d_idx] = pln[p_idx];
				}else{
					data[data_idx].real = 0.0f;
					data[data_idx].imag = 0.0f;
				}
			}
		}
	}

	#pragma offload target(mic:0))\
		nocopy(prop : length(N2*N2) alloc_if(0) free_if(0))
	{
		//Transfer function
		#pragma omp parallel for
		for(int j=0;j<N2;j++){
			for(int i=0;i<N2;i++){
				unsigned int idx = i + N2 * j;
				float dx;
				float dy;

				if(x < N && y < N){
					dx = (float)(i);
					dy = (float)(j);
				}else if(x >= N && y < N){
					dx = (float)(x - N2);
					dy = (float)(y);
				}else if(x < N && y >= N){
					dx = (float)(x);
					dy = (float)(y - N2);
				}else{
					dx = (float)(x - N2);
					dy = (float)(y - N2);
				}

				float th = PI*P*(dx*dx+dy*dy)/(D*WL);
				prop.real = cosf(th);
				prop.imag = sinf(th);
			}
		}
	}

	//create and initialize DFTI descriptor
	#pragma offload target(mic:0) nocopy(hand0,hand1)
	{
		MKL_LONG NN[2];	NN[0] = N2;	NN[1] = N2;
		//create DFTI descriptor
		DftiCreateDescriptor(&hand0 ,DFTI_SINGLE ,DFTI_COMPLEX, 2, NN);
		DftiCreateDescriptor(&hand1 ,DFTI_SINGLE ,DFTI_COMPLEX, 2, NN);
		//initialize
		DftiCommitDescriptor(hand0);
		DftiCommitDescriptor(hand1);
	}

	//data plane and propagation term FFT Forward
	#pragma offload target(mic:0) nocopy(hand0,hand1)\
		nocopy(data : length(N2*N2) alloc_if(0) free_if(0))\
		nocopy(prop : length(N2*N2) alloc_if(0) free_if(0))\
	{
		DftiComputeForward(hand0, data);
		DftiComputeForward(hand1, prop);
	}

	//complex multiplication
	#pragma offload target(mic:0)\
		nocopy(data : length(N2*N2) alloc_if(0) free_if(0))\
		nocopy(prop : length(N2*N2) alloc_if(0) free_if(0))\
	{
		#pragma omp parallel for
		for(int i=0;i<N2*N2;i++){
			float dre = data[i].real;
			float dim = data[i].imag;
			float pre = prop[i].real;
			float pim = prop[i].imag;
			data[i].real = dre*pre-dim*pim;
			data[i].imag = dre*pim+dim*pre;
		}
	}

	//FFT Backward
	#pragma offload target(mic:0) nocopy(hand0)\
		nocopy(data : length(N2*N2) alloc_if(0) free_if(0))
	{
		DftiComputeBackward(hand0,data);
	}

	//data cripping and trasfer result from Xeon Phi to host
	#pragma offload target(mic:0)\
		nocopy(data : length(N2*N2) alloc_if(0) free_if(0)\
		out(pln : length(N*N) alloc_if(0) free_if(0)
	{
		#pragma omp parallel for
		for(int j=0;j<N;j++){
			for(int i=0;i<N;i++){
				int d_idx = (i + N/2) + (j + N/2) * N2;
				int p_idx = i + j * N;
				pln[idx] = d[idx];
			}
		}
	}

	//memory and descriptor free
	#pragma offload target(mic:0)\
		nocopy(hand0,hand1)\
		nocopy(pln : length(N2*N2) alloc_if(0) free_if(1))\
		nocopy(data : length(N2*N2) alloc_if(0) free_if(1))\
		nocopy(prop : length(N2*N2) alloc_if(0) free_if(1))
	{
		DftiFreeDescriptor(&hand0);
		DftiFreeDescriptor(&hand1);
	}
}
\end{lstlisting}

We describe how to use FFT using Intel MKL simply.
\begin{enumerate}
\item declare DFTI handler (in line 17,18 of List \ref{lst:fresnel_CPU}, \\the type of handler is ``DFTI\_DESCRIPTOR\_HANDLE").
\item create DFTI descriptor (in line 74,75 of List \ref{lst:fresnel_CPU}, ``DftiCreateDescriptor" function) and initialize the descriptor (in line 77,78 of List \ref{lst:fresnel_CPU}, ``DftiCommitDescriptor" function).
\item call FFT computation function (in line 81,82 of List \ref{lst:fresnel_CPU}, ``DftiComputeForward" function computes forward FFT, in line 96, ``DftiComputeBackward" function computes inverse FFT).
\item free DFTI descriptor (in line 109,110 of List \ref{lst:fresnel_CPU}, ``DftiFreeDescriptor" function).
\end{enumerate}

In line 28-41 of List \ref{lst:fresnel_CPU}, the area of source plane is expanded from $N \times N$ to $2N \times 2N$ during calculation because of avoiding aliasing by circular convolution. 
After the calculation, we need to clip the wanted area whose size is $N \times N$ (in line 99-106 of List \ref{lst:fresnel_CPU}).

We explain pragma directives and option keywords in List \ref{lst:fresnel_Phi}.
In lines 31 and 32 of List \ref{lst:fresnel_Phi}, ``nocopy" is a keyword that indicates the data is not transferred between the host and Xeon Phi coprocesser.
The modifier ``alloc\_if()" indicates allocating memory explicitly on the Xeon Phi coprocessor, ``alloc\_if(1)" means allocating memory, ``alloc\_if(0)" means not allocating memory.
Similarly, the modifier ``free\_if()" indicates freeing memory explicitly on the Xeon Phi coprocessor.

In contrast, List \ref{lst:fresnel_GPU} shows the implementation on the GPU by CUDA. 
We used CUFFT library  provided by NVIDIA as the FFT library.
CUFFT library has an interface similar to the FFTW library that is widely used in the world. 
Comparing Lists \ref{lst:fresnel_Phi} and \ref{lst:fresnel_GPU}, the GPU programming is troublesome because kernel functions must be written for expanding area, clipping area, calculating the impulse response of Eq.(\ref{eqn:fresnel}) and complex multiplication.
Furthermore, memory allocations and kernel execution using CUDA APIs are required.
%
\begin{lstlisting}[caption={Source code on GPU by CUDA (Fresnel diffraction).}, label=lst:fresnel_GPU, language=C, numbers=left, numberstyle=\tiny, stepnumber=1, frame=single]
#define N	1024				//size N*N of input plane			
#define N2	2048				//2*N

#define P	8.0e-6F				//sampling pitch
#define WL	633e-9				//wavelength
#define PI	3.1415926535F

#define D	1.0					//propagated distance

#define TX 32					//number of threads (x direction)
#define TY 32					//number of threads (y direction)

//the area is expanded N*N to N2*N2
__global__ void Expand(cufftComplex *pln, cufftComplex *data){
	int x = threadIdx.x + blockIdx.x * blockDim.x;
	int y = threadIdx.y + blockIdx.y * blockDim.y;

	unsigned int N_h = N >> 1;
	unsigned int dx = x + N_h;
	unsigned int dy = y + N_h;

	unsigned int idx = x + N2 * y;
	unsigned int didx = dx + N * dy;

	data[didx] = pln[idx];
}

//data cripping N2*N2 to N*N
__global__ void Crip(cufftComplex *pln, cufftComplex *data){
	int x = threadIdx.x + blockIdx.x * blockDim.x;
	int y = threadIdx.y + blockIdx.y * blockDim.y;

	unsigned int N_h = N >> 1;
	unsigned int dx = x + N_h;
	unsigned int dy = y + N_h;

	unsigned int idx = x + N2 * y;
	unsigned int didx = dx + N * dy;

	pln[idx] = data[didx];
}

//Transfer function
__global__ void PropTerm(cufftComplex *prop){
	int x = threadIdx.x + blockIdx.x * blockDim.x;
	int y = threadIdx.y + blockIdx.y * blockDim.y;

	unsigned int idx = x + N2 * y;
	
	float dx;
	float dy;
	
	if(x < N && y < N){
		dx = (float)(i);
		dy = (float)(j);
	}else if(x >= N && y < N){
		dx = (float)(x - N2);
		dy = (float)(y);
	}else if(x < N && y >= N){
		dx = (float)(x);
		dy = (float)(y - N2);
	}else{
		dx = (float)(x - N2);
		dy = (float)(y - N2);
	}

	float th = PI*P*(dx*dx+dy*dy)/(D*WL);
	prop[idx] = make_cuComplex(cosf(th),sinf(th));
}

//complex multiplication
__global__ void ComplexMult(cufftComplex *data, cufftComplex *prop){
	int x = threadIdx.x + blockIdx.x * blockDim.x;
	int y = threadIdx.y + blockIdx.y * blockDim.y;
	
	unsigned int idx = x + N2 * y;

	data[idx] = cuCmulf(data[idx],prop[idx]);
}

int main(void){
	
	cufftComplex *pln;			//pointer for source and destination plane array (for CPU)
	cufftComplex *d_pln;		//pointer for source and destination plane array (for GPU)
	cufftComplex *d_data;		//data(expanded source and destination area) array (for GPU)
	cufftComplex *d_prop;		//pointer for propagation term array (for GPU)
	
	//host memory allocation		
	pln = (cufftComplex*)malloc(sizeof(cufftComplex)*N2*N2);
	
	//GPU memory allocation
	cudaMalloc((void**)&d_pln,sizeof(cufftComplex)*N*N);
	cudaMalloc((void**)&d_data,sizeof(cufftComplex)*N2*N2);
	cudaMalloc((void**)&d_prop,sizeof(cufftComplex)*N2*N2);

	//transfer plane data from host to GPU
	cudaMemcpy(d_pln,pln,sizeof(cufftComplex)*N*N,cudaMemcpyHostToDevice);
	cudaMemset(d_data,0.0f,sizeof(cufftComplex)*N2*N2);
	
	dim3 ex_blocks(N/TX,N/TY,1);
	dim3 ex_threads(TX,TY,1);

	Expand<<<ex_blocks,ex_threads>>>(d_pln,d_data);

	dim3 blocks(N2/TX,N2/TY,1);
	dim3 threads(TX,TY,1);

	//propagation term 
	PropTerm<<<blocks,threads>>>(d_prop);

	cufftHandle plan;
	cufftPlan2d(&plan,N2,N2,CUFFT_C2C);

	//data plane and propagation term FFT
	cufftExecC2C(plan,d_prop,d_prop,CUFFT_FORWARD);
	cufftExecC2C(plan,d_data,d_data,CUFFT_FORWARD);

	//complex multiplication
	ComplexMult<<<blocks,threads>>>(d_data,d_prop);

	//data plane IFFT
	cufftExecC2C(plan,d_data,d_data,CUFFT_INVERSE);

	//data cripping
	Crip<<<ex_blocks,ex_threads>>>(d_pln,d_data);
	
	//trasfer result from GPU to host
	cudaMemcpy(pln,d_pln,sizeof(cufftComplex)*N*N,cudaMemcpyDeviceToHost);

	cufftDestroy(plan);
	cudaFree(d_pln);
	cudaFree(d_data);
	cudaFree(d_prop);
}
\end{lstlisting}

\section{Results}
In this section, we compare the CGH computational times of the direct calculation, recurrence relation algorithm, Fresnel diffraction and wavefront recording method on the Xeon Phi coprocessor, Multi-core CPU, and GPU.

\subsection{Hardware environment}
Table \ref{tab:hardware_environment} shows the hardware environment in this work.
The CPUs we used are two Xeon E5-2643 CPUs. The GPU we used is NVIDIA Tesla K20. Table \ref{tab:k20_spec} shows the specification of NVIDIA Tesla K20 GPU.
Afterward, two Xeon E5-2643 are called the CPUs, Xeon Phi 5110P is called the Xeon Phi, and NVIDIA Tesla K20 is called the GPU.

\begin{table}[htbp]
  \begin{center}
    \begin{tabular}{cc}

      \begin{minipage}{0.6\hsize}
        \begin{center}
			\caption{Hardware environment}
			\label{tab:hardware_environment}
			\begin{tabular}{c|c} \hline
				OS & CentOS 6.4 \\
				CPU & Intel Xeon E5-2643 (4 Core) $\times$ 2 \\ 
				Xeon Phi & Intel Xeon Phi 5110P\\
				GPU & NVIDIA Tesla K20\\
				Compiler & Intel Compiler XE 13.1 \\
				GPU Compiler & CUDA 5.0 \\
				\hline
			\end{tabular}
        \end{center}
      \end{minipage}

      \begin{minipage}{0.4\hsize}
        \begin{center}
          \caption{Specification of NVIDIA Tesla K20 (GPU).}
          \label{tab:k20_spec}
          \begin{tabular}{c|c} \hline
			Core & 2496 \\
			Clock & 0.706 [GHz] \\
			Memory & 5 [GB] \\
			Memory Band Width & 208 [GB/s] \\ \hline
          \end{tabular}
        \end{center}
      \end{minipage}

    \end{tabular}
  \end{center}
\end{table}

\subsection{Transfer time to host PC}
Table \ref{tab:transfer_time} shows the transfer times to the host PC by the Xeon Phi and GPU. The data size is approximately 2M bytes ($1,920 \times 1,080 [\mathrm{pixels}] \times 8 [\mathrm{bytes / pixel}]$), which is typical CGH size.
As shown in Table \ref{tab:transfer_time}, the Xeon Phi transfers the data 1.9 times faster than the GPU.
\begin{table}[h]
	\caption{Transfer time to the host PC.}
	\label{tab:transfer_time}
	\begin{center}
	\begin{tabular}{c|c} \hline
	 Processor & transfer time [$\mu$s] \\ \hline
	 Xeon Phi & 330 \\ 
	 GPU & 630  \\ \hline
	\end{tabular}
	\end{center}
\end{table}

\subsection{Comparison of performance}
Table \ref{tab:result_direct} shows the CGH computational times of the direct calculation.
Table \ref{tab:result_recurrence} also shows the CGH computational times of the recurrence relation method.
We used the CGH size of $1,920 \times 1,080$ pixels.
The pixel has the single-precision floating point number.
We used the following compile options for each processor:
\begin{itemize}
\item ``icc test.c -fast -openmp -mkl -fp-model fast=2'' (for the CPU)
\item ``icc test.c -fast -openmp -mkl -fp-model fast=2'' (for the Xeon Phi)
\item ``nvcc test.cu -O3 -gencode arch=compute\_35, -code=sm\_35, -use\_fast\_math -lcufft'' (for the GPU)
\end{itemize}
where the intel compiler (icc) option ``-fast'' means that the whole program is compiled at the maximum speed (this option is composed of ``-ipo,'' which is the inlining and other interprocedural optimizations among multiple source files, ``-O3,'' which is the maximum optimization, ``-no-prec-div,'' which is use of the fast and low-precision division function, ``-static,'' which is the static linker option, and ``-xhost,'' which is generating instruction sets supported by the compilation host), ``-mkl'' means using Intel MKL, ``-fp-model fast=2'' means the aggressive optimization on floating-point data (but precision may be slightly low), ``-gencode arch=compute\_35 -code=sm\_35'' means the GPU compute capability is 3.5, ``-use\_fast\_math'' means using the fast and low-precision math library, and ``-lcufft'' means using the CUFFT library. All of the calculations on the CPUs uses the full 16 CPU threads.

As shown in Table \ref{tab:result_direct}, in the direct calculation of CGH, the Xeon Phi computed about 8 times faster than the CPUs when the number of object points is over 10,000. 
In contrast, the GPU is 100 times faster than the CPUs and 12 times faster than the Xeon Phi.
As shown in Table \ref{tab:result_recurrence}, in the recurrence relation method, the Xeon Phi is slower than using the direct calculation. 
One of the reason for this is that the sequential part of the CGH calculation in the recurrence relation method is increased, as compared with the direct calculation.

\begin{table}[htbp]
    
	\begin{minipage}[c]{0.5\hsize}
		\begin{center}
			\caption{Computational time[sec] \\ (direct calculation).}
			\label{tab:result_direct}
			\begin{tabular}{c||c|c|c} \hline
				\multicolumn{1}{c||}{} & \multicolumn{3}{c}{Number of object points} \\  
								&	1,000	&	10,000	& 100,000 \\ \hline 
				 CPU			&	1.7		&	16.3		& 163 \\
				 Xeon Phi		&	1.02		&	2.01		& 20.2 \\
				 GPU			&	0.016	&	0.168	& 1.68 \\ \hline
			\end{tabular}
		\end{center}
		
	\end{minipage}
	\hfill
	\begin{minipage}[c]{0.5\hsize}
		
		\begin{center}
			\caption{Computational time[sec] \\ (recurrence relation method).}
			\label{tab:result_recurrence}
			\begin{tabular}{c||c|c|c} \hline
				\multicolumn{1}{c||}{} & \multicolumn{3}{c}{Number of object points} \\  
								&	1,000	&	10,000	&	100,000	\\ \hline 
				 CPU			&	0.92		&	8.23		&	83.3			\\
				 Xeon Phi		&	1.46		&	7.26		&	42.1			\\
				 GPU			&	0.014	&	0.141	&	1.51		\\ \hline
			\end{tabular}
		\end{center}
		
	\end{minipage}
\end{table}

Table \ref{tab:result_diffraction} shows the calculation times for the Fresnel diffraction. The size of the computational plane is $N^2 = (2^m)^2$ (m is a natural number) that can calculate FFT efficiently.
When the size of the computational plane is $8,192 \times 8,192$, the GPU could not calculate it because of insufficient memory.
The Xeon Phi and the GPU computed respectively about 2.4 and 16.5 times faster than the CPUs in the case of $N \times N=4,096 \times 4,096$. 

\begin{table}[h]
		\begin{center}
			\caption{Computational time[sec] \\ (Fresnel diffraction).}
			\label{tab:result_diffraction}
			\begin{tabular}{c||c|c|c|c} \hline
				\multicolumn{1}{c||}{} & \multicolumn{4}{c}{Size of plane $N \times N$} \\  
				  				&	$1,024 \times 1,024$	&	$2,048 \times 2,048$	&	$4,096 \times 4,096$	&	$8,192 \times 8,192$	\\ \hline 
				 CPU			&	0.17			&	0.60		&	2.06		&	9.00		\\
				 Xeon Phi		&	0.32			& 	0.41		&	0.87		&	3.80		\\
				 GPU			&	0.011		&	0.027	&	0.125	&  N.A.	\\ \hline
			\end{tabular}
		\end{center}
\end{table}

\begin{table}[h]
	\begin{center}
		\caption{Computational time[sec] \\ (wavefront recording method).}
		\label{tab:result_wrp}
		\begin{tabular}{c||c|c|c} \hline
			\multicolumn{1}{c||}{} & \multicolumn{3}{c}{Number of object points} \\  
							&	1,000	&	10,000	& 100,000	\\ \hline 
			 CPU			&	1.02		&	3.04		& 9.34		\\
			 Xeon Phi		&	0.68		&	0.70		& 0.72		\\
			 GPU			&	0.029	&	0.030	& 0.035		\\ \hline
		\end{tabular}
	\end{center}
\end{table}

\begin{figure}[h]
	\begin{center}
		\includegraphics[width=100mm]{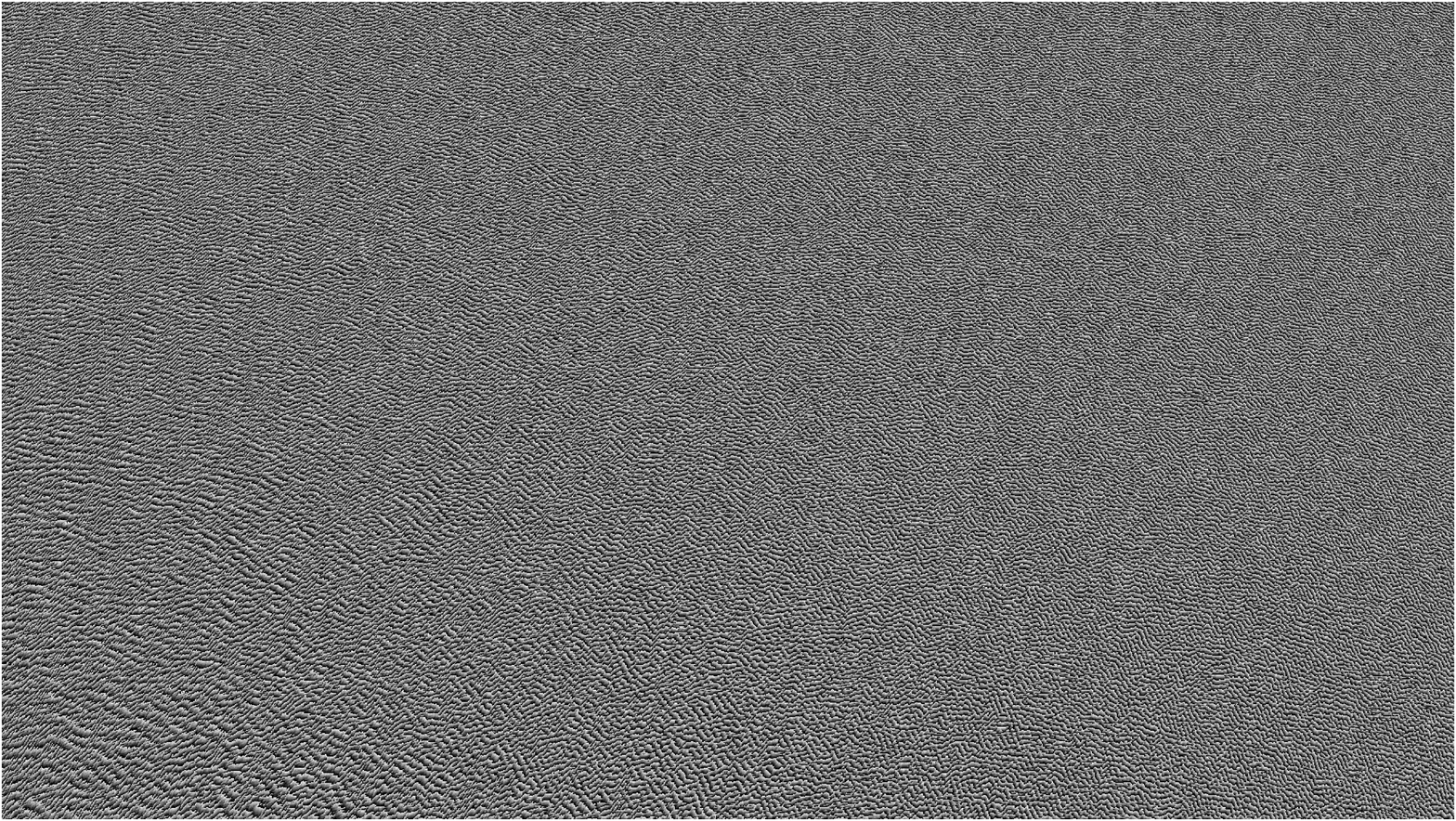}
	\end{center}
	\caption{CGH calculated on the Xeon Phi coprocessor.}
	\label{fig:CGH}
\end{figure}

\begin{figure}[h]
	\begin{center}
		\includegraphics[width=100mm]{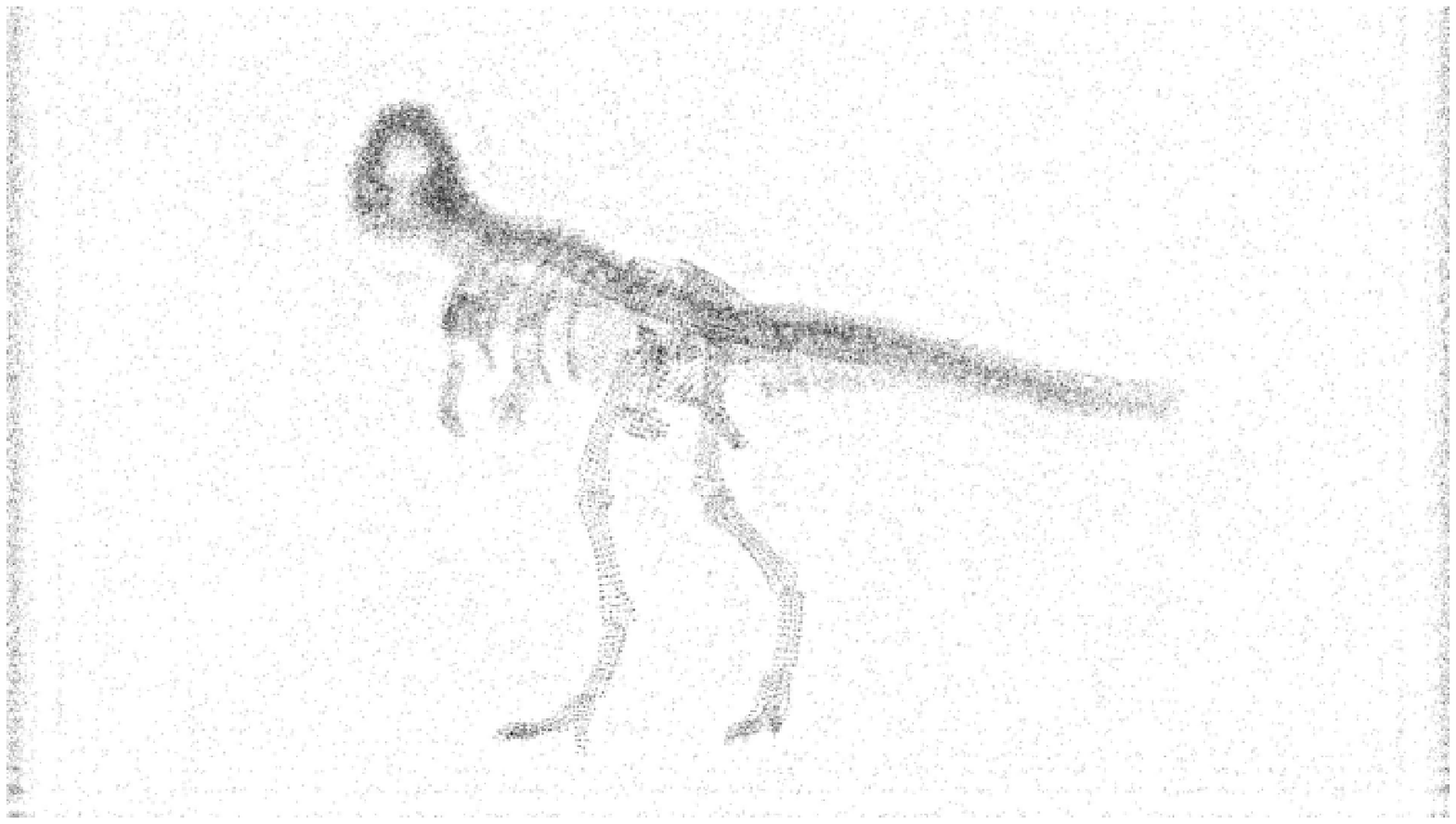}
	\end{center}
	\caption{Reconstructed image (simulation).}
	\label{fig:reconstructed_image}
\end{figure}

Table \ref{tab:result_wrp} shows the CGH computational times of wavefront recording method. We optimized the first step on the GPU \cite{wr3}.
Figures \ref{fig:CGH} and \ref{fig:reconstructed_image} show the CGH calculated on the Xeon Phi and the reconstructed image, respectively. 
The calculation condition is as follows: The CGH size is $1,920 \times 1,080$ [pixels]. The number of object points is 11,646. The wavelength is 531.8 [nm].

\section{Conclusion}
We implemented the CGH computation on the Xeon Phi coprocessor by adding only a few offload pragma directives to the CPU source code, which is much easier compared with GPU programming.
As a result, in the direct calculation, we succeeded in achieving  calculation about 8 times faster than the CPUs when we used the full 16 CPU threads, and about 13-times faster than the CPUs when using the wavefront recording method.
We also succeeded in achieving the Fresnel diffraction calculation two-times faster than the CPUs. 

In all cases, the GPU is faster than the Xeon Phi coprocessor. However, if we want to implement an existing code for a CPU on a GPU, it is necessary to rewrite many parts on the existing source code for the GPU.
In contrast, in the Xeon Phi coprocessor, it is unnecessary to rewrite many parts of the existing source code by adding only a few pragma directives. 
Therefore, we can reduce the working hours for the implementation and try various algorithms readily.

\section*{Acknowledgments}
\noindent This work is supported by Japan Society for the Promotion of Science (JSPS) KAKENHI (Grant-in-Aid for Scientific Research (C) 25330125) 2013, and KAKENHI (Grant-in-Aid for Scientific Research (A) 25240015) 2013.

\end{document}
\endinput